\documentclass{ws-ijmpd}
\begin{document}

\markboth{B.J. Ahmedov and F.~J.~Fattoyev} {Quasi-stationary
Electromagnetic Effects  in Schwarzschild Space-time}

%
\catchline{}{}{}{}{}
%

\title{QUASI-STATIONARY ELECTROMAGNETIC EFFECTS IN CONDUCTORS\\
AND SUPERCONDUCTORS IN SCHWARZCHILD SPACE-TIME}
\author{\footnotesize B. J. AHMEDOV$^\ast$ and F. J. FATTOYEV$^\dag$}
\address{Institute of Nuclear Physics and
    Ulugh Beg Astronomical Institute, Astronomicheskaya 33,
    Tashkent 700052, Uzbekistan\\
    and\\
IUCAA, Post Bag 4 Ganeshkhind,
411007 Pune, India\\
   $^\ast$ahmedov@astrin.uzsci.net\\
   $^\dag$farrooh@iucaa.ernet.in}
\maketitle



\begin{abstract}
The general principles needed to compute the effect of a
stationary gravitational field on the quasistationary
electromagnetic phenomena in normal conductors and superconductors
are formulated from general relativistic point of view.
Generalization of the skin effect, that is the general
relativistic modification of the penetration depth (of the
time-dependent magnetic field in the conductor) due to its
relativistic coupling to the gravitational field is obtained. The
effect of the gravitational field on the penetration and coherence
depths in superconductors is also studied. As an illustration of
the foregoing general results, we discuss their application to
superconducting systems in the outer core of neutron stars. The
relevance of these effects to electrodynamics of magnetized
neutron stars has been shown.

\end{abstract}





\section{Introduction}

Since electromagnetic system is a purely relativistic system, its
coupling to gravity can be described, in principle, only
general-relativistically. In our previous research, the effect of
the gravitational field on stationary electromagnetic processes in
conductors and superconductors in the presence of constant
electric current has been investigated. In fact, the influence of
angular momentum of the gravitational source may appear as a
galvanogravitomagnetic effect in current carrying
conductors,\cite{a99} as general-relativistic effect of charge
distribution inside conductors in an applied magnetic
field,\cite{ak00} as general-relativistic effects in charged
systems.\cite{ar03} The effect of gravitational field on
thermoelectric effects in superconductors and on electromagnetic
properties of type II superconductors have been investigated in
our papers.\cite{a99a,ak04} Here we extend our results to the
time-dependent electromagnetic processes.

The paper is organized as follows. First we write Maxwell
equations in the Schwarzschild space-time in Section~\ref{meq}.
Then in the next part of the paper, we will study the effect of
gravity on the quasistationary time-varying electric current and
magnetic field penetrating the conductor placed in the
gravitational field. Sections~\ref{skin_schw}
is devoted to investigation of the skin effect in the
Schwarzschild space-time and in a weak gravitational field when
the curvature effects are locally negligible. In particular,  new
purely general-relativistic effects connected with the
gravitational modification of penetration depth of magnetic field
will be obtained.

The remarkable quantum phenomena exhibited by a superconductor
show that the superconducting state in Ginzburg-Landau
theory\cite{gl50} described by a wavefunction can be spread over
the entire superconductor. The existence of macroscopic
wavefunction raises the natural question of how it is influenced
by the gravitational field. In the last part of the paper, that
is, in Sections~\ref{coherence_length}-\ref{penet_NS} we shall
formulate the general formulae and their solutions needed to
determine the influence of stationary gravitational field on the
quasistationary electromagnetic phenomena for superconductors,
from  general-relativistic point of view and study the relevance
of general-relativistic effects in superconductors of II type on
electrodynamics of magnetized neutron stars. This analysis has
been motivated by the mixture commonly found in neutron star
models, namely in the outer core region, where superfluid
neutrons, superconducting protons and normal electrons are
generally thought to coexist (see for example, Ref. 8). However,
due to the generality of the present approach, it is equally well
applicable to superconducting systems found in more common
laboratory contexts in the weak gravitational field of Earth.

In this paper we use a space-like signature $(-,+,+,+)$ and a
system of units in which $G = 1 = c$ (however, for those
expressions with a physical application we have written the speed
of light explicitly). Greek indices are taken to run from 0 to 3
and Latin indices from 1 to 3; covariant derivatives are denoted
with a semi-colon and partial derivatives with a comma.

\section{Maxwell Equations in Schwarzschild Space-time}
\label{meq}

In a coordinate system $(t,r,\theta,\phi)$, the metric for a
static spherically-symmetric relativistic star with the total mass
$M$ is
\begin{equation}
\label{schw} ds^2 = - N^2 dt^2 +N^{-2} dr^2 + r^2 d\theta ^2+
r^2\sin^2\theta d\phi^2 \ ,
\end{equation}
where  $N^2\equiv1-{2M}/{r}$ is the lapse function.

    The general form of the first pair of general
relativistic Maxwell equations is given by
\begin{equation}
\label{maxwell_firstpair}
F_{\alpha \beta, \gamma }
    + F_{\gamma \alpha, \beta} + F_{\beta \gamma,\alpha} = 0 \ ,
\end{equation}
where $F_{\alpha \beta}$ is the electromagnetic field tensor
expressing the strict connection between the electric $E^{\alpha}$
and magnetic $B^{\alpha}$ fields. For an observer with
four-velocity $(u^{\alpha})_{_{\rm obs}}$, the covariant
components of the electromagnetic tensor are given by
\begin{equation}
\label{fab_def} F_{\alpha\beta} \equiv 2 (u)_{_{\rm obs}[\alpha}
E_{\beta]} +
    \eta_{\alpha\beta\gamma\delta}(u^{\gamma})_{_{\rm obs}} B^\delta \ ,
\end{equation}
where $A_{[\alpha \beta]} \equiv \frac{1}{2}(A_{\alpha \beta} -
A_{\beta \alpha})$ and $\eta_{\alpha\beta\gamma\delta}$ is the
pseudo-tensorial expression for the Levi-Civita symbol
$\epsilon_{\alpha \beta \gamma \delta}$
\begin{equation}
\eta^{\alpha\beta\gamma\delta}=-\frac{1}{\sqrt{-g}}
    \epsilon_{\alpha\beta\gamma\delta} \ ,
    \hskip 2.0cm
\eta_{\alpha\beta\gamma\delta}=
    \sqrt{-g}\epsilon_{\alpha\beta\gamma\delta} \ ,
\end{equation}
with $g\equiv {\rm
det}|g_{\alpha\beta}|=- r^4\sin^2\theta$ for the metric (\ref{schw}).

The proper observers have
four-velocity components given by
\begin{equation}
\label{uprops}
(u^{\alpha})_{_{\rm prop}}\equiv
    N^{-1}\bigg(1,0,0,0\bigg) \ ;
    \hskip 2.0cm
(u_{\alpha})_{_{\rm prop}}\equiv
    N\bigg(- 1,0,0,0 \bigg) \ .
\end{equation}

    In the metric (\ref{schw}) and
with the definition (\ref{fab_def}) referred to the observers
(\ref{uprops}), the first pair of Maxwell equations
(\ref{maxwell_firstpair}) take then the form\cite{ra04}
\begin{eqnarray}
\label{max1a}
&& \sin\theta \left(r^2B^{\hat r}\right)_{,r}+
    \frac{r}{N}\left(\sin\theta B^{\hat \theta}\right)_{,\theta} +
    \frac{r}{N} B^{\hat \phi}_{\ , \phi} = 0 \ , \\
\label{max1b} &&
\left({\frac{r}{N}\sin\theta}\right)\frac{\partial B^{\hat
r}}{\partial t}
  = E^{\hat\theta}_{\ ,\phi}- \left(\sin\theta
    E^{\hat \phi} \right)_{,\theta} \ ,
\\
\label{max1c}
&& \left(\frac{r}{N}\sin\theta\right)
    \frac{\partial B^{\hat \theta}}{\partial t}
    =\sin\theta \left(r N E^{\hat \phi} \right)_{,r}
    -E^{\hat r}_{ ,\phi}\ ,
\\
\label{max1d}
&& \left(\frac{r}{N}\right)
    \frac{\partial B^{\hat \phi}}{\partial t}
    = E^{\hat r}_{\ ,\theta}-\left(r N E^{\hat \theta}\right)_{,r} \ ,
    \end{eqnarray}
where "hatted" quantities are projected onto a locally orthonormal
tetrad $\{{{\bf e}}_{\hat \mu}\} = ({\bf e}_{\hat 0}, {\bf
e}_{\hat r}, {\bf e}_{\hat \theta}, {\bf e}_{\hat \phi})$ carried
by a proper observer
\begin{equation}
\label{prop_tetrad} {\bf e}_{\hat 0}^{\alpha}  =
    \frac{1}{N}\bigg(1,0,0,0\bigg) \ ,\qquad
    {\bf e}_{\hat r}^{\alpha}  =
    N \bigg(0,1,0,0\bigg) \ ,\qquad
    {\bf e}_{\hat \theta}^{\alpha}  =
    \frac{1}{r}\bigg(0,0,1,0\bigg)  \ ,   \qquad
    {\bf e}_{\hat \phi}^{\alpha}  =
    \frac{1}{r\sin\theta}\bigg(0,0,0,1\bigg) \ .
\end{equation}

    The general form of the second pair of Maxwell
equations is given by
\begin{equation}
\label{maxwell_secondpair} F^{\alpha \beta}_{\ \ \ \ ;\beta} =
4\pi J^{\alpha} \,
\end{equation}
where the four-current $J^{\alpha}$ is a sum of convection and
conduction $j^{\alpha}$ currents
\begin{equation}
J^{\alpha}=\rho_e u^\alpha + j^\alpha \ ,
    \hskip 2.0cm j^\alpha u_\alpha \equiv 0 \ ,
\end{equation}
with $u^{\alpha}$ being the conductor four-velocity as whole and
$\rho_e$ the proper charge density.

We can now rewrite the second pair of Maxwell equations
as\cite{ra04}
\begin{eqnarray}
\label{max2a}
&& \sin\theta\left(r^2 E^{\hat r} \right)_{,r}+
    \frac{r}{N} \left(\sin\theta E^{\hat \theta}\right)_{,\theta}
    + \frac{r}{N} E^{\hat \phi}_{\;,\phi}
     =\frac{4\pi r^2\sin\theta}{N} J^{\hat t}\ ,
\\
\label{max2b} && \left(\sin\theta  B^{\hat \phi} \right)_{,\theta}
    - B^{\hat\theta}_{\ ,\phi}
    = \left({\frac{r\sin\theta}{N}}\right)
    \frac{\partial E^{\hat r}}{\partial t}
    +{4\pi}r\sin\theta J^{\hat r} \ ,
\\
\label{max2c}
&& B^{\hat r}_{\ ,\phi} - \sin\theta \left(r N
    B^{\hat \phi} \right)_{,r}
    =  \left(\frac{r\sin\theta}{N}\right)
    \frac{\partial E^{\hat\theta}}{\partial t}
    +{4\pi}r\sin\theta J^{\hat\theta} \ ,
\\
\label{max2d}
&& \left(r N B^{\hat \theta} \right)_{,r}
 - B^{\hat r}_{\ ,\theta}
 = \left(\frac{r}{N}\right)
    \frac{\partial E^{\hat\phi}}{\partial t}
    +{4\pi} rJ^{\hat\phi} \ .
\end{eqnarray}

\section{Skin Effect in Schwarzschild Gravitational Field}
\label{skin_schw}

If the conduction current $j^\alpha$ is carried by the
electrons with
electrical conductivity $\sigma$, Ohm's law can then be
written as
\begin{equation}
\label{ohm} j_\alpha = \sigma F_{ \alpha \beta}u^\beta \ .
\end{equation}

    Equations (\ref{max2b})--(\ref{max2d}) can now be
rewritten in a more convenient form. Taking four-velocity
components of the conductor as (\ref{uprops}), we can use Ohm's
law (\ref{ohm}) to derive the following explicit components of
current $J^{\hat \alpha}$ in the proper frame
\begin{equation}
\label{current1} J^{\hat t} = {\rho_e} \ , \hskip 2.0cm
 J^{\hat i}=\sigma E^{\hat i} \ .
\end{equation}

\noindent Hereafter we will use the following realistic
assumptions. Firstly, we consider $\sigma$ to be uniform within
the conducting media. Secondly, we ignore the contributions coming
from the displacement currents. The latter could, in principle, be
relevant in the evolution of the electromagnetic fields, but their
effects are negligible on timescales that are long as compared
with the electromagnetic waves crossing time. In view of this, we
will neglect in (\ref{max2b})--(\ref{max2d}) all terms involving
time derivatives of the electric field (quasistationarity
condition) and use Ohm's law (\ref{current1}) to rewrite Maxwell
equations (\ref{max2b}) and (\ref{max2c}) as
\begin{eqnarray}
\label{rel1}
&& E^{\hat r} =
    \frac{c}{4\pi\sigma r\sin\theta}\left[
    \left(\sin\theta B^{\hat \phi}\right)_{\;,\theta}
    - B^{\hat\theta}_{\; ,\phi}\right] \ ,
\\
\label{rel2}
&& E^{\hat\theta} =
    \frac{c}{4\pi\sigma r\sin\theta}\left[
    B^{\hat r}_{\ ,\phi}- \sin\theta
    \left(r N B^{\hat \phi} \right)_{,r}\right] \ ,
\\
\label{rel3}
&& E^{\hat\phi} =
     \frac{c}{4\pi\sigma r}\left[
     \left(r N B^{\hat \theta}\right)_{,r}
     -B^{\hat r}_{\ ,\theta}\right] \ .
\end{eqnarray}

Using Maxwell equation (\ref{max2a}) and Ohm's law
(\ref{current1}), we find that the space charge density $\rho_e =
\rho_e(t, r, \theta, \phi)$ inside the star has a zeroth-order
contribution given by
\begin{eqnarray}
\label{rho}
&& \rho_e = \frac{c N}{4\pi r^2\sin\theta}
    \left[\sin\theta \left(r^2 E^{\hat r} \right)_{,r}+
    \frac{r}{N} \left(\sin\theta
 E^{\hat\theta} \right)_{,\theta}+
 \frac{r}{N} E^{\hat\phi}_{,\phi} \right] \ .
    \end{eqnarray}

Now with the help of Eqs. (\ref{max1b})--(\ref{max1d}),
(\ref{rel1})--(\ref{rel3}) and Ohm's laws (\ref{current1}), we
obtain the induction equations for the components of time varying
magnetic field penetrating the conductor

\begin{eqnarray}
\label{ind1}
&& \frac{\partial B^{\hat r}}{\partial t} =
    \frac{c^2 N}{4\pi\sigma r^2\sin\theta}
    \Bigg\{\frac{1}{\sin\theta}
    \left[B^{\hat r}_{\ ,\phi} -
    \sin\theta \left(r N B^{\hat \phi} \right)_{,r}
    \right]_{,\phi}
    -\left(\sin\theta
    \left[\left(r N B^{\hat \theta} \right)_{,r}
    - B^{\hat r}_{\ ,\theta}\right]\right)_{,\theta}\Bigg\} \ ,
 \\ \nonumber\\
\label{ind2}
&& \frac{\partial B^{\hat \theta}}{\partial t} =
    \frac{c^2 N}{4\pi\sigma r\sin\theta}
    \Bigg\{\sin\theta \left(N \left[\left(r N
    B^{\hat \theta}\right)_{,r}
    -B^{\hat r}_{, \theta} \right]\right)_{,r}
    -\frac{1}{r\sin\theta}
    \left[ \left(\sin\theta
    B^{\hat \phi} \right)_{,\theta}-B^{\hat \theta}_{\ ,\phi}
    \right]_{,\phi}\Bigg\}  \ ,
 \\ \nonumber\\
\label{ind3}
&& \frac{\partial B^{\hat\phi}}{\partial t} =
    \frac{c^2 N}{4\pi\sigma r}\Bigg\{\left(
    \frac{1}{r \sin\theta}\left[\left(\sin\theta
    B^{\hat \phi}\right)_{,\theta}-B^{\hat\theta}
    _{, \phi}\right]\right)_{, \theta}
    -\left(\frac{N}{\sin\theta}\left[B^{\hat r}_{,\phi}-
    \sin\theta \left(r N B^{\hat \phi} \right)_{,r} \right]
    \right)_{,r}\Bigg\}
  \ .
 \end{eqnarray}

Assume a conducting sphere with the conductivity $\sigma$ occupies
the space $r\leq R$ with empty vacuum space in the region $r>R$.
The surface at $r=R$ is subjected to dipolar magnetic field
varying harmonically with time. For example, the azimuthal
component of magnetic field $B^{\hat\theta}$ is
\begin{equation}
\label{skin1}  B^{\hat\theta} = B^{\hat\theta}_{0}e^{-i\omega
 t} \ .
\end{equation}

In order to understand the behavior of the magnetic field inside
the spherical conductor we look for a solution of the induction
equation (\ref{ind2}) in the inner region $r\leq R$
\begin{equation}
\label{solution3} -i \omega B^{\hat\theta} = \frac{c^2 N}
{4\pi\sigma r}\left(N \left[r N B^{\hat\theta}\right]_{,r}
\right)_{,r} \ ,
\end{equation}
which can be written as the following differential expression
 \begin{equation}
 \label{key1}
  \frac{\partial^2 B^{\hat \theta}}{\partial r^2}+
 \left(\frac{3M}{r^2N^2}+\frac{2}{r}\right)
 \frac{\partial B^{\hat \theta}}{\partial r}+k^2 B^{\hat \theta}
 +\frac{M}{r^3N^2} B^{\hat\theta} = 0  \ ,
 \end{equation}
 where the parameter $k^2=(4\pi\sigma i \omega)/(c^2N^3)$ is introduced.

 Analytical solution of this equation can not be found easily, and
 as we do not need any numerical solution for our purpose we must
 choose some approximation that do not destroy the structure
 of possible solution.
 Due to this purpose we will see that in linear $M/r$
 approximation this equation could be simplified and
 written as Bessel equation
 \begin{equation}
 \label{key2}
 \frac{\partial^2 B^{\hat \theta}}{\partial r^2}+
 \frac{2}{r}\frac{\partial B^{\hat \theta}}
 {\partial r}+k^2 B^{\hat \theta} = 0\ .
 \end{equation}
 The boundary condition for this equation is
 $B_0^{\hat\theta}=B^{\hat\theta}(R)$. And the solution of this
 equation is the spherical
 Bessel function of the first type\cite{arfken}
 \begin{equation}
 \label{key3}
 B^{\hat\theta}=B^{\hat\theta}_{0}\frac{j_{0} (kr)}{j_{0}(kR)} \ ,
 \end{equation}
 where
    \begin{equation}
    \label{delta}
 k =\frac{1+i}{\delta} \ ,\qquad \delta= \frac{cN^{3/2}}{\sqrt{2\pi\sigma\omega}}  \ .
\end{equation}

We will expand the expression for the magnetic field~(\ref{key3})
at the surface of the sphere for the large arguments of $kr\gg1$.
If we take $x=R-r$ as a distance from the surface inwards then the
interior magnetic field behaves as
\begin{equation}
 \label{key4}
 B^{\hat\theta}= B^{\hat\theta}_{0}e^{ikx}=B^{\hat\theta}_{0}
 e^{i\frac{x}{\delta}}+B^{\hat\theta}_{0}e^{-\frac{x}{\delta}} \ ,
 \end{equation}
and the distance $x=\delta$ is called the skin depth of magnetic
field penetration (see Fig. 1).
\begin{figure}
\centerline{\psfig{figure=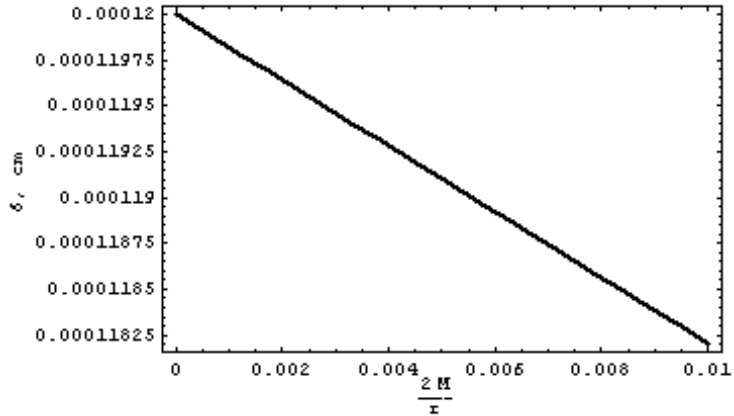,width=3.8in}}
\caption{The dependence of skin depth in the proton stellar
superconductor on the compactness parameter $\frac{2M}{r}$ in the
case when conductivity $\sigma\approx 10^{25} s^{-1}$ and the
frequency $\omega\approx 10^3 s^{-1}$.}
\end{figure}

 From Eq. (\ref{rel3}) one can find the expression
\begin{equation}
\label{rel4}
E^{\hat\phi}=\frac{ckN}{4\pi\sigma}\frac{j_1(kr)}{j_0(kR)}B^{\hat\theta}_0
\ ,
\end{equation}
for azimuthal component of the electric field.

Thus the electric field falls off exponentially in $r$, with a
spatial oscillation of the same scale, being confined mainly to a
depth less than the relativistically modified skin depth.

\subsection{Skin effect in weak Earth's gravitational field}
\label{skin_earth}

In the weak gravitational field of the Earth the curvature effects
are negligible and space-time metric in the Cartezian coordinates
is
 \begin{equation}
 \label{earth_metric}
 ds^2=-c^2N^2dt^2+dx^2+dy^2+N^{-2}dz^2     \ .
\end{equation}
 Here $N$ is
equal to
         \begin{equation}
         \label{N2}
   N=1+\frac{2gz}{c^2}\ ,
     \end{equation}
and $z$ is the height above some fixed point, when the apparatus
at the Earth may be regarded as having an acceleration $g$
relative to local inertial frame.

    The evolution equations for the magnetic field in the metric (\ref{earth_metric})
     take the form
    \begin{eqnarray}
    \label{mag.f.1}
    && \frac{\partial B^{\hat x}}{\partial
    t}=\frac{c^2}{4\pi\sigma}N
    \Bigg\{\left[N\left[\left(N B^{\hat x}\right)_{,z}-B^{\hat z}_{,x}
    \right]\right]_{,z}- \left[B^{\hat y}_{,x}-
    B^{\hat x}_{,y}\right]_{,y}\Bigg\} \ ,
    \\ \nonumber\\
    \label{mag.f.2}
    && \frac{\partial B^{\hat y}}{\partial t}=
    \frac{c^2}{4\pi\sigma}N \Bigg\{\left[B^{\hat
    y}_{,x}-B^{\hat x}_{,y}\right]_{,x}-
     \left[N\left[B^{\hat   z}_{,y}-\left(N B^{\hat
    y}\right)_{,z}\right]\right]_{,z}\Bigg\}  \ ,
    \\ \nonumber\\
    \label{mag.f.3}
    && \frac{\partial B^{\hat z}}{\partial t}=
    \frac{c^2}{4\pi\sigma}N \Bigg\{\left[B^{\hat z}_{,y}-   \left(N
  B^{\hat y}\right)_{,z}\right]_{,y}-
    \left[\left(N B^{\hat x}\right)_{,z}
    -B^{\hat z}_{,x}\right]_{,x}\Bigg\}  \ .
      \end{eqnarray}

Assume a semi-infinite conductor of uniform conductivity $\sigma$
occupying the space $z>0$. The surface at $z=0$ is subjected to
tangential magnetic field $B^{\hat x}=B^{\hat x}_0e^{-i\omega t}$
governed by the differential equation
\begin{equation}
\label{earth's.skin1}
\frac{\partial^2 B^{\hat x}}{\partial z^2}+
\frac{6g}{c^2 N}\frac{\partial B^{\hat x}}{\partial z}+
\frac{4g^2}{c^4 N^2}B^{\hat x} +k^2 B^{\hat x}=0\ ,
\end{equation}
which has a solution
 \begin{equation}
 \label{earth's.skin2}
  B^{\hat x}=C_1e^{{(-3c^2gN-\sqrt{5c^4g^2N^2-c^8k^2N^4})z}/{c^4N^2}}
  +C_2e^{{(-3c^2gN+\sqrt{5c^4g^2N^2-c^8k^2N^4})z}/{c^4N^2}}\ ,
\end{equation}
where $C_1$ and $C_2$ are the integration constants. Neglecting
small terms and using (\ref{key4}) for $k$ one can get
\begin{equation}
\label{earth's.skin4}
  B^{\hat x}=B^{\hat x}_0e^{iz/{\delta}}e^{-z/{\delta}}\ .
  \end{equation}

\section{Effect of Gravity on Coherence Length and
Penetration Depth} \label{coherence_length}

    The coherence length is the measure of the distance within
which the properties of the superconductors are not changed
appreciably in the presence of a magnetic field. Because we are
concerned with the effect of gravity on the coherence length in a
superconductor it would be useful first to discuss Ginzburg-Landau
equation in a gravitational field\cite{gl50}
\begin{equation}
\label{g-l1} \alpha\psi+ \beta\psi{|\psi|}^2+
\frac{1}{4m}{\left(i\nabla+2eA\right)}^2\psi=0\ ,
\end{equation}
 which could be written as
 \begin{equation}
 \label{g-l2}
 \alpha\psi+\beta\psi{|\psi|}^2+
\frac{1}{4m}g^{\mu\nu}{\left(-{\nabla}_{\mu}{\nabla}_{\nu} +
2ei{\nabla}_{\mu}A_{\nu}+2eiA_{\mu}{\nabla}_{\nu}+4e^2{A}_{\mu}A_{\nu}\right)}\psi=0\
.
\end{equation}
Here $\psi$ is the complex order parameter describing the
superconductor, $\alpha$ and $\beta$ are the phenomenological
expansion coefficients.

We now rewrite Ginzburg-Landau equation (\ref{g-l1}) in the
Schwarzschild metric (\ref{schw}) in a more useful form
\begin{eqnarray}
\label{g-l3}
&& \alpha\psi+ \beta\psi{|\psi|}^2
\nonumber\\
&& +\frac{1}{4m}\Bigg\{-\left[-\frac{1}
{N^2}\frac{{\partial}^2}{\partial
t^2}+\frac{1}{r^2\sin\theta}\left(
\sin\theta\frac{\partial}{\partial
r}\left[r^2N^2\frac{\partial}{\partial r}\right]+
\frac{\partial}{\partial \theta}\left[\sin\theta
\frac{\partial}{\partial
\theta}\right]
+\frac{1}{\sin\theta}\frac{{\partial}^2}{\partial
{\phi}^2} \right)\right]
\nonumber\\
&& + 2ei\left[-\frac{1}{N^2}\frac{\partial A_{t}}{\partial t}+
\frac{1}{r^2\sin\theta}\left(\sin\theta\frac{\partial}{\partial r}
\left[r^2N^2A_{r}\right] +
\frac{\partial}{\partial\theta}\left[\sin\theta A_{\theta}\right]+
\frac{1}{\sin\theta}\frac{\partial A_{\phi}}{\partial
\phi}\right)\right]
\nonumber\\
&& + 2ei\left[-\frac{A_t}{N^2}\frac{\partial}{\partial t}+
\frac{A_r}{r^2\sin\theta}\left(\sin\theta\frac{\partial}{\partial
r} \left[r^2N^2\right] +
A_{\theta}\frac{\partial}{\partial\theta}\left[\sin\theta \right]+
\frac{A_{\phi}}{\sin\theta}\frac{\partial }{\partial
\phi}\right)\right]
\nonumber\\
&& + 4e^2\left[-\frac{1}{N^2}A_{t}^2 +N^2 A_{r}^2 +\frac{1}{r^2}
A_{\theta}^2
+\frac{1}{r^2{\sin}^2\theta}A_{\phi}^2\right]\Bigg\}\psi=0\ .
\end{eqnarray}

 Assuming that the order parameter $\psi$ has no variation in coordinates
 $\theta$ and $\phi$  and all components of vector-potential of electromagnetic field
 vanish then Eq. (\ref{g-l3}) reduces to
\begin{equation}
\label{g-l4} \alpha\psi+\beta\psi{|\psi|}^2-\frac{1}{4m}N^2\left(
\frac{{\partial}^2\psi}{\partial
r^2}+\frac{2}{r}\left[1+\frac{M}{rN^2}\right]\frac{\partial\psi}
{\partial r}\right) =0\ ,
\end{equation}
which in approximation $M/r\ll1$ takes the simple form
\begin{equation}
\label{g-l5} \alpha\psi+\beta\psi{|\psi|}^2-\frac{1}{4m}N^2
\left(\frac{{\partial}^2\psi} {\partial
r^2}+\frac{2}{r}\frac{\partial\psi}{\partial r}\right)=0 \ .
\end{equation}

Now we introduce the dimensionless wavefunction $\varphi(r)$
\begin{equation}
\label{unsized}
 \varphi(r)=\frac{\psi(r)}{\psi_0}\ ,
\end{equation}
where
\begin{equation}
\label{psi}
 {\psi}_0^2(r)=\frac{n_s}{2}=\frac{|\alpha|}{\beta}\ ,
\end{equation}
\begin{equation}
\label{g-l6} -\varphi+\varphi{|\varphi|}^2-{\xi}^2\left(
\frac{{\partial}^2\varphi}{\partial r^2}
+\frac{2}{r}\frac{\partial\varphi}{\partial r}\right)=0\ .
\end{equation}

    Here the parameter ${\xi}^2= {\xi}_0^2N^2=\frac{N^2}{4m{|\alpha|}}$,
    $\xi_0$ is the Ginzburg-Landau coherence length in the flat
    Minkowski space-time, $n_{(s)}$ is the density of superconducting electrons.
     The coherence length is the length scale on which
    the condensate charge density $2e|\psi|^2$ vanishes, as one
    approaches the surface of the superconductor from its
    interior.

     Now we distinguish the meaning of $\xi$.
The coordinate axis $r$ is perpendicular to the surface of the
superconductor $r=R$. Assume that the surface layer is slim and
the value of the function $\varphi$ at the surface has small
deflection from $1$, that is
 \begin{equation}
 \label{phi}
  \varphi=1-\epsilon(r)\ ,
 \end{equation}
 and substituting the expression for $\varphi$ into (\ref{g-l6})
 we obtain the Bessel equation\cite{arfken}
    \begin{equation}
    \label{epsilon}
  \frac{{\partial}^2\epsilon}{\partial r^2}+\frac{2}{r}
    \frac{\partial\epsilon}{\partial r}-2{\xi}^{-2}\epsilon=0
    \end{equation}
 in the linear approximation in $\epsilon$.

     Then taking into account the property $\lim_{r \to 0}\varphi=1$
     one can get $\epsilon(0)=0$ and find the solution of the  Eq. (\ref{epsilon}) as modified
    Bessel function of the first type\cite{arfken}
    \begin{equation}
    \label{epsilon2}
    \epsilon=\epsilon(0)\frac{j_{0}(i\sqrt{2}\frac{r}{\xi})}{j_{0}(i\sqrt{2}\frac{R}{\xi})} \ .
    \end{equation}

It is known that the spherical Bessel function $j_0(r)$ of zeroth
order for large arguments $r\gg1$ is approximated as
\begin{equation}
\label{BESSELapp}
j_0(r)=\frac{\sin{r}}{r}\ ,
\end{equation}
and due to this approximation near the surface of the sphere
expression (\ref{epsilon2}) takes the form
 \begin{equation}
    \label{epsilon3}
    \epsilon=\epsilon(0)e^{-\sqrt{2}\frac{x}{\xi}} \ ,
    \end{equation}
where $x=R-r$ is the inward distance from the surface of the
sphere. It follows that $\xi$ is the length scale of the parameter
of order $\varphi$ (See Fig. 2).
\begin{figure}
\centerline{\psfig{figure=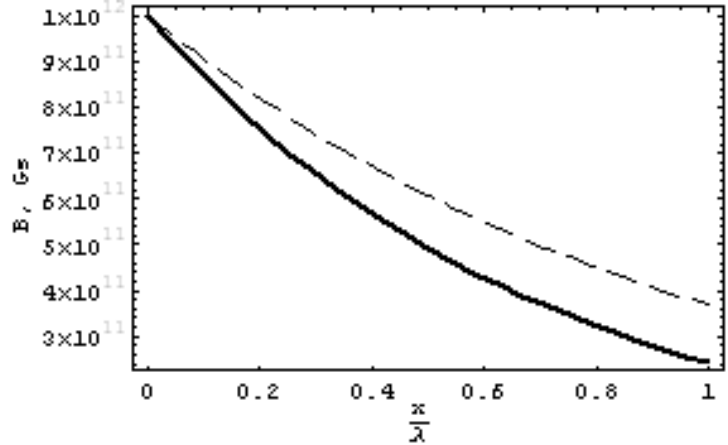,width=3.8in}}
\caption{Exponential growth of the order parameter with the
distance $x=R-r$ from the surface of the gravitational object
toward its interior; normal line is responsible for Newtonian flat
space-time and dashed line is for neutron star with the
compactness parameter ${2M}/{r}=0.5$.}
\end{figure}

    In the relativistically generalized London theory, the equation for
the superconducting current $j^{\alpha}_{s}$ is\cite{a97}
\begin{equation}
\label{london} 2j_{(s)[\alpha;\beta]}-2j_{(s)[\alpha}(\ln
n_{(s)})_{,\beta]}
+c\rho_{0(s)}(A_{\beta\alpha}+2u_{[\alpha}w_{\beta]})=
\frac{c}{4\pi}{\lambda}_L^{-2}F_{\alpha\beta}\ ,
\end{equation}
where the notation ${\lambda}_L^2=(mc^2)/(4\pi n_{(s)}e^2)$ for
the London penetration depth is introduced and
\begin{equation}
\label{relrot}
A_{\alpha\beta}=u_{[\alpha,\beta]}+u_{[\beta}w_{\alpha]}\
\end{equation}
is the relativistic rate of rotation,   $w_{\alpha}=u_{\alpha,
\beta}u^{\beta}$ is the absolute acceleration of superconductor.

    Using the London equation (\ref{london}) we obtain the expressions
for the electric field (we choose c=1 for this case too)
\begin{eqnarray}
\label{electr1}
&& E^{\hat r}= 4\pi{\lambda}_L^2[N^{-1}j^{\hat
r}_{,t}- (N j^{\hat t})_{,r}-N^{-1} j^{\hat r}(\ln n)_{,t}+ N
j^{\hat t} (\ln n)_{,r}+ \rho N_{,r}] \  ,
 \\ \nonumber\\
\label{electr2}
&& E^{\hat \theta}=
4\pi{\lambda}_L^2[N^{-1}j^{\hat\theta}_{,t}- r^{-1}j^{\hat
t}_{,\theta}-N^{-1}j^{\hat\theta}(\ln n)_{,t}+ r^{-1}j^{\hat
t}(\ln n)_{,\theta}]   \ ,
\\ \nonumber\\
\label{electr3}
 && E^{\hat\phi}=4\pi{\lambda}_L^2[N^{-1}j^{\hat
\phi}_{,t}- r^{-1}{\sin}^{-1}\theta j^{\hat t}_{,\phi}-
N^{-1}j^{\hat\phi}(\ln n)_{,t} +r^{-1}{\sin}^{-1}\theta j^{\hat
t}(\ln n)_{,\phi}]\ .
\end{eqnarray}

Here the derivatives from the density $n_{(s)}$ of superconducting
pairs are taken to be equal to zero, because the dependence of
$n_{(s)}$ on the coordinates is comparably small. Putting $j^{\hat
i}$ extracted from the second pair of the Maxwell equations
(\ref{max2b})--(\ref{max2d}) into expressions
(\ref{electr1})--(\ref{electr3}) in the framework of this
approximation and inserting the obtained result into the first
pair of Maxwell equations (\ref{max1a})--(\ref{max1d}) one obtains
the expressions for magnetic field which are the second London
equations
\begin{eqnarray}
\label{londoneqn1} &&B^{\hat
r}_{,t}=\frac{{\lambda}_L^2}{r\sin\theta}\Bigg\{\frac{1}
{r\sin\theta}\left(B^{\hat r}_{,\phi}-\sin\theta\left[rNB^{\hat
\phi}\right]_{,r} \right)_{,t\phi}
-\left(\frac{\sin\theta}{r}\left(\left[rNB^{\hat\theta}\right]_{,r}
-B^{\hat r}_{,\theta}\right)_{,t}\right)_{,\theta}\Bigg\} \ ,
\\ \nonumber\\
\label{londoneqn2}
&&B^{\hat\theta}_{,t}=\frac{{\lambda}_L^2N}{r\sin\theta}
\Bigg\{\sin\theta\left(\left[rNB^{\hat\theta}\right]_{,r}-B^{\hat
r}_{,\theta} \right)_{,tr}-\frac{1}{Nr\sin\theta}
\left(\left[\sin\theta B^{\hat\phi}\right]_{,\theta}
-B^{\hat\theta}_{,\phi}\right)_{,t\phi}\Bigg\} \ ,
\\ \nonumber\\
\label{londoneqn3} &&B^{\hat
\phi}_{,t}=\frac{{\lambda}_L^2N}{r}\Bigg\{\left(\frac{1}{Nr\sin\theta}
\left(\left[\sin\theta
B^{\hat\phi}\right]_{,\theta}-B^{\hat\theta}_{,\phi}\right)
\right)_{,t\theta} -\frac{1}{\sin\theta}\left(B^{\hat
r}_{,\phi}-\sin\theta
\left[rNB^{\hat\phi}\right]_{,r}\right)_{,tr}\Bigg\} \ .
\end{eqnarray}
For simplicity of calculations the London equation
(\ref{londoneqn2})
  for $B^{\hat\theta}$
component of the magnetic field
\begin{equation}
\label{pendepth1}
B^{\hat\theta}=\frac{{\lambda}_L^2N}{r}\left(rNB^{\hat\theta}
\right)_{,rr}=
{\lambda}_L^2N^2\left(-\frac{M^2}{r^4N^4}B^{\hat\theta}+
\frac{2}{r}B^{\hat\theta}_{,r}+\frac{2M}{r^2N^2}
B^{\hat\theta}_{,r}+B^{\hat\theta}_{,rr}\right)\ ,
\end{equation}
can be approximately written as Bessel like equation
\begin{equation}
\label{pendepth2}
B^{\hat\theta}_{,rr}+\frac{2}{r}B^{\hat\theta}_{,r}-
{\lambda}^{-2}B^{\hat\theta}=0\ ,
\end{equation}
in the linear approximation in the small compactness parameter
$M/r$, where parameter $\lambda={\lambda}_LN$.

The solution of the Eq. (\ref{pendepth2}) is already known and
expressed through the Bessel function
\begin{equation}
\label{pendepth3} B^{\hat\theta}=
B^{\hat\theta}_0\frac{j_{0}(i\frac{r}{\lambda})}{j_{0}(i\frac{R}{\lambda})}
\ .
\end{equation}
\begin{figure}
\centerline{\psfig{figure=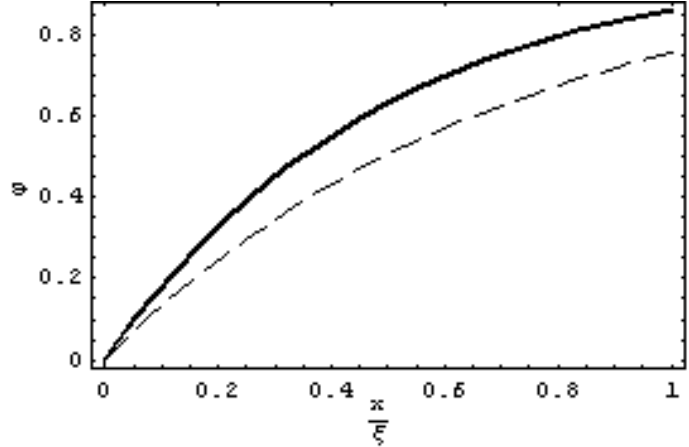,width=3.8in}}
\caption{The exponential decay of magnetic strength inside a
superconductor of type II, normal line is for flat space-time,
dash line is for a neutron star with the compactness parameter
$\frac{2M}{r}=0.5$. Here $x=R-r$ is the distance from the surface
of the star toward its interior.}
\end{figure}

Using the above mentioned arguments one may get the following
expression at the surface of the sphere (see Fig. 3)
\begin{equation}
\label{pendepth4} B^{\hat\theta}=
B^{\hat\theta}_0e^{-\frac{x}{\lambda}} \ .
\end{equation}

Our main finding is the dependence of the coherence length and
penetration depth on the gravitational field. It is connected with
the general relativistic modification of surface electromagnetic
energy of the conductors and superconductors due to the effect of
gravitational field.

\section{Magnetic Field of Single Vortex}
\label{vortex2}

The four-vector of supercurrent density is\cite{a99a}
\begin{equation}
\label{g-liia}
 j_{(s)\alpha}=-\frac{ie}{m_s}
 \Bigg\{\psi^*(\partial_{\alpha}-2eiA_{\alpha})\psi
-\psi(\partial_{\alpha}+2eiA_{\alpha})\psi^{*}\Bigg\} \ ,
\end{equation}
which could be written through  dimensionless function
$\varphi=\psi/\psi_0$  as
 \begin{equation}
 \label{g-liib}
j_{(s)\alpha}=-i\frac{\Phi_0}{(4\pi\lambda)^2}
\left(\varphi^*\partial_{\alpha}\varphi-
\varphi\partial_{\alpha}\varphi^*\right)-
\frac{{|\varphi|}^2}{4\pi\lambda^2}A_{\alpha} \ ,
\end{equation}
where
\begin{equation}
\label{Phi_0}  \Phi_0=\frac{hc}{2e}  \
  \end{equation}
is the quantum of magnetic flux and
\begin{equation}
 \label{lambda}
 \lambda^2=\frac{m\beta c^2}{8\pi e^2 |\alpha|}   \ .
\end{equation}
Assume that the wavefunction is $\varphi=|\varphi|e^{i\vartheta}$.
Then superconducting current density (\ref{g-liib}) takes the form
\begin{equation}
\label{g-liic}
 j_{(s)\alpha}=\frac{{|\varphi|}^2}{4\pi\lambda^2}\left(
 \frac{\Phi_0}{2\pi}\partial_{\alpha}\vartheta-A_{\alpha}
 \right)\approx
 \frac{1}{4\pi\lambda^2}\left(\frac{\Phi_0}{2\pi}
 \partial_{\alpha}\vartheta-A_{\alpha}\right)   \ .
 \end{equation}
 If the parameter of the Ginzburg-Landau
 theory $\kappa=\lambda/\xi$ is much bigger than $1$ then at large distance
 $r\gg\xi$ we will have $|\varphi|^2\approx 1$.

 In cylindrical coordinates $(t, r, \phi, z)$ the space-time metric
 around a gravitating source takes the form\cite{ll71}
\begin{equation}
 \label{cylindr}
 ds^2=-N^2dt^2+N^{-2}(dr^2+r^2d\phi^2+dz^2)  \ .
 \end{equation}
Using this metric we take curl from both sides of Eq.
(\ref{g-liic}) and obtain the differential equations for the
components of magnetic field  $B^{\hat i}$:
\begin{eqnarray}
 \label{g-liida}
 &&\frac{1}{r}\left[
    (rB^{\hat \phi})_{,r}-B^{\hat r}_{,\phi}\right]_{,\phi}-
    r\left[B^{\hat r}_{,zz}
    -B^{\hat z}_{,rz}\right]
 =\frac{r}{\lambda^2 N^2}\Bigg\{\frac{\Phi_0}
    {2\pi}\left({\mathbf{\nabla}}\times\nabla\vartheta\right)^{\hat r}
-B^{\hat r}\Bigg\} \ ,
    \\ \nonumber\\
    \label{g-liidb}
 &&\frac{1}{r}\left[B^{\hat z}_{,\phi z}- rB^{\hat\phi}
     _{,zz}\right]-\left(\frac{1}{r}\left[(rB^{\hat\phi})_{,r}-
     B^{\hat r}_{,\phi}\right]\right)_{,r}
 =\frac{1}{\lambda^2 N^2}\Bigg\{\frac{\Phi_0}{2\pi}
     \left({\mathbf{\nabla}}\times\nabla\vartheta\right)^{\hat\phi}
-\frac{1}{r}B^{\hat\phi}\Bigg\} \ ,
     \\ \nonumber\\
    \label{g-liidc}
 &&\left(r\left[B^{\hat r}_{,z}
    -B^{\hat z}_{,r}\right]\right)_{,r}-
    \frac{1}{r}\left [B^{\hat z}_{,\phi\phi}-rB^{\hat\phi}_{,z\phi}
    \right]=\frac{r}{\lambda^2 N^2}\Bigg\{\frac{\Phi_0}{2\pi}
    \left({\mathbf{\nabla}}\times\nabla\vartheta\right)^{\hat z}-B^{\hat z}\Bigg\}\ .
     \end{eqnarray}

According to Stokes Theorem\cite{misner}
 \begin{equation}
 \label{vortex}
 ({\mathbf{\nabla}}\times\nabla\vartheta)^{\mu} =2\pi
\delta(x^{\alpha}- x^{\alpha}_0){\eta}^{\mu} \ ,
 \end{equation}
here $\eta^{\mu}$ is the unit vector aligned along the vortex.

Assume $B^{\hat\phi}=B^{\hat r}=0$ and $B^{\hat z}(r)\neq 0$, then
Eq. (\ref{g-liidc}) could be written as equation for the $z$
component of magnetic field
\begin{equation}
\label{g-liidb2}
 B^{\hat z}-\lambda^2N^2\left(B^{\hat z}_{,rr}+
\frac{1}{r}B^{\hat z}_{,r}\right)= \Phi_0\delta(r){e}^{\hat z} \ ,
\end{equation}
where $e^{\hat z}$ is the unit vector along $z$ axis. This is the
Bessel equation of complex argument with the boundary condition
$B^{\hat z}(\infty)=0$. The solution of the Eq. (\ref{g-liidb2})
is the Bessel function of the third type, either the McDonald
function or Hankel function of complex argument
\begin{equation}
\label{hankel}  B^{\hat z}=\frac{\Phi_0}{2\pi\lambda^2 N^2}
K_0(r/\lambda N) \ .
\end{equation}
The asymptotic behaviour of Hankel function of complex argument
leads to the following result
\begin{eqnarray}
\label{hankel2a} && B^{\hat z}=\frac{\Phi_0}{2\pi\lambda^2
N^2}\left(\frac{\pi\lambda N}{2r}\right)^{1/2}e^{-\frac{r}{\lambda
N}} ,\qquad for \qquad r\gg\lambda N \,
\\ \nonumber\\
 \label{hankel2b}
&& B^{\hat z}=\frac{\Phi_0}{2\pi\lambda^2
N^2}\ln\left(\frac{\lambda N}{r}\right),\qquad \hskip 1.48 cm for
\qquad r\ll\lambda N \ .
\end{eqnarray}

It follows  from (\ref{hankel}) and (\ref{hankel2b}) that in the
centre of vortex the magnetic field tends to infinity. However, in
reality the field can not be infinite and these formulae are not
valid near normal core of vortex ($r\sim\xi$). Then the expression
(\ref{hankel2b}) can be written as
\begin{equation}
 \label{B0}
 B(0)=\frac{\Phi_0}{2\pi\lambda^2 N^2}\ln\kappa \ .
 \end{equation}

\section{Penetration of Magnetic Field into II Type Superconductor}
\label{generalization}

Assume that space-time admits a time-like Killing vector
$\xi^\alpha$
\begin{equation}
\label{killing1} \xi^\alpha\xi_\alpha=-\Lambda \ ,
\end{equation}
where $-\Lambda=-N^2$ is the norm of time-like Killing vector,
$u^\alpha$ is the four-velocity of superconductor as a whole being
parallel to the time-like
 Killing vector $\xi^\alpha$
 \begin{equation}
 \label{4-vel1}
 u^\alpha u_\alpha=-1 \ , \hskip 2.0cm
 \xi^\alpha=Nu^\alpha\ .
 \end{equation}
 Then 4-momentum $P^\alpha$ of the superconducting Cooper pairs is
 \begin{equation}
 \label{4-moment}
 P^\alpha=mcu^\alpha_s\ .
 \end{equation}
 The four-velocity  of superconducting electrons $u^\alpha_s$
  can be decomposed in the form
    \begin{equation}
    \label{4-vel2}
    u^\alpha_s=\frac{u^\alpha+v^\alpha_s}{\sqrt{1-v^2_s}} \ ,
    \end{equation}
 where $v^\alpha_s$ is the relative velocity of the Cooper pairs.

The energy of a superconducting pair is
    \begin{equation}
    \label{energy1}
     W=-P_\alpha\xi^\alpha=mcu^\alpha_s\xi_\alpha=
    -mc\frac{u^\alpha\xi_\alpha}{\sqrt{1-v^2/c^2}}=
    N\frac{mc^2}{\sqrt{1-v^2/c^2}}\ .
    \end{equation}
In the case of nonrelativistic motion, the kinetic energy of
superconducting pair is
    \begin{equation}
    \label{energy2}
    W=Nmc^2+N\frac{mv_s^2}{2} \ .
 \end{equation}
The density of total kinetic energy of superconducting pair is
\begin{equation}
    \label{energy4}
    W_{kin}=N\frac{mj_s^2}{2n_se^2}\ .
    \end{equation}
    where $j_s^\alpha=n_sev_s^\alpha$ is the superconducting current
    density.

Taking into account Maxwell equations written in vector form
    \begin{equation}
    \label{maxwella}
    \mathbf{\nabla}\times(N\vec{B})=-\frac{4\pi}{c}\vec{j_s} \,
    \end{equation}
one can get the expression for kinetic energy (\ref{energy4}) in a
more useful form
    \begin{equation}
    \label{energy5}
    W_{kin}=N\frac{\lambda^2}{8\pi}(\mathbf{\nabla}\times(N\vec{B}))^2 \ .
    \end{equation}

 According to expressions for the kinetic energy of supercurrent
 (\ref{energy4}) and for energy of magnetic field the free energy of
 the whole superconductor is equal to
 \begin{equation}
 \label{frenergy}
 {\mathcal{F}}_{\mathrm{sB}}={\mathcal{F}}_{\mathrm{s0}}+
  \frac{1}{8\pi}N\int[\vec{B}^2+\lambda^2(\mathbf{\nabla}\times(N\vec{B}))^2]\sqrt{-g}dV \ .
 \end{equation}

 According to Leibnitz formula
 \begin{equation}
 \label{math}
 \left(\eta^{\alpha\beta\mu\nu}u_{\beta}A_{\mu}B_{\nu}\right)_{;\alpha}=
 \left(\eta^{\alpha\beta\mu\nu}u_{\beta}A_{\mu;\alpha}\right)B_{\nu}+
 \left(\eta^{\alpha\beta\mu\nu}u_{\beta}B_{\nu;\alpha}\right)A_{\mu} \ ,
 \end{equation}
Expressing vector potential $A_{\mu}$ through the magnetic field
\begin{equation}
\label{math2}
A_{\mu}=g_{\mu\sigma}\eta^{\sigma\gamma\tau\lambda}u_{\gamma}B_{\tau;\lambda}
\,
\end{equation}
and using Stokes Theorem\cite{misner}
\begin{equation}
\label{gauss}
\int\left(\eta^{\alpha\beta\mu\nu}u_{\beta}A_{\mu}B_{\nu}\right)_{;\alpha}
\sqrt{-g}dV=\oint
\left(\eta^{\alpha\beta\mu\nu}u_{\beta}A_{\mu}B_{\nu}\right)n_{\alpha}dS
= 0 \,
\end{equation}
one could obtain the following expression for full energy of
superconductor
\begin{equation}
\label{frenergy2} {\mathcal{E}}=\frac{N}{8\pi} \int
\vec{B}(\vec{B}+\lambda^2\mathbf{\nabla \times \nabla \times}(N
\vec{B}))\sqrt{-g}dV \ .
\end{equation}
Taking into account that magnetic field $\vec{B}$ is governed by
Eq. (\ref{g-liidb2}) one can get
\begin{equation}
\label{fullenergy} {\mathcal{E}}=\frac{\Phi_0N}{8\pi}B(0) \ .
\end{equation}
Inserting (\ref{B0}) into (\ref{fullenergy}) one can get
\begin{equation}
\label{fullenergy2}
 {\mathcal{E}}=\left(\frac{\Phi_0}{4\pi\lambda
N}\right)^2 N\ln\kappa \ .
\end{equation}

For the superconductor embedded in applied external magnetic field
the density of Gibbs free energy for the unit length of vortex
\begin{equation}
\label{gibbs} \mathcal{G}={\mathcal{E}}-N\frac{\Phi_0B_0}{4\pi} \
.
\end{equation}
takes the minimum value at the equilibrium.

From this formula one can see that for the comparatively weak
field $B_0$, Gibbs energy $\textsl{G}>0$ and the formation of
vortex is impossible. However there exists the magnetic field
$B_{c1}$ from which $\textsl{G}$ becomes negative and vortex
formation becomes energetically favorable. It follows from
(\ref{gibbs}) that
\begin{equation} \label{Bc1}
B_{c1}=\frac{4\pi\mathcal{E}}{N\Phi_0}=
\frac{\Phi_0}{4\pi\lambda^2N^2}\ln\kappa\ .
\end{equation}
The inner magnetic field should be found as a solution of the
interior Maxwell equations. However for the superconductor
embedded in an applied magnetic field $B_{out}$, the interior
magnetic field can be modelled as $B_0=B_{out}N^{-2}$ for
simplicity of calculations. Therefore, the free energy for the
superconductor is
\begin{equation}
\label{frenergy2b}
F_{sB}=F_{s0}+\frac{B_{out}^2}{8\pi}N=F_{s0}+\frac{B_0^2}{8\pi}N^5
\ .
\end{equation}
When $F_{sB}=F_n$, the superconductor is transformed to the normal
state and
 \begin{equation}
 \label{normalstate}
 F_n-F_{s0}=\frac{B_{cm}^2}{8\pi}N^5  \ .
 \end{equation}

 On the other hand according to Ginzburg-Landau theory\cite{gl50}
 the free energy of superconductor
 in the gravitational field in the absence of  external magnetic
 field is
\begin{equation}
\label{frenergy3}
F_{s0}=F_n+N\alpha{|\psi|}^2+N\frac{\beta}{2}{|\psi|}^4  \ .
\end{equation}

One can find the value of ${|\psi|}^2$ in which free energy has a
minimum, by solving $\frac{dF_{s0}}{d{|\psi|}^2}=0$. Simple
calculation gives
\begin{equation}
\label{psisquare}
{|\psi|}^2=-\alpha/\beta   \ .
\end{equation}

Inserting (\ref{psisquare}) to (\ref{frenergy3}) and equalizing
the result to (\ref{normalstate}) we will have
\begin{equation}
\label{Bcm}
 B_{cm}^2=\frac{4\pi\alpha^2}{\beta N^4}  \ .
\end{equation}
Then using the expressions for $\xi$, $\lambda$ and $\Phi_0$ we
obtain for $B_{cm}$
\begin{equation}
\label{Bcm2}
 B_{cm}=\frac{\sqrt{2}\Phi_0}{4\pi\lambda\xi N^2}  \ .
\end{equation}
The second critical magnetic field\cite{a57} is
\begin{equation}
\label{Bc2} B_{c2}=\sqrt{2}\kappa B_{cm}=\frac{\Phi_0}{2\pi\xi^2
N^2}  \ .
\end{equation}

\section{Application of General Relativistic Electromagnetic Effects
to Neutron Stars Interior}
\label{penet_NS}

The dependence of the coherence length and penetration depth on
the gravitational field may be important for neutron star physics.
There are two ways in which external or trapped magnetic flux can
penetrate the proton superfluid of neutron stars. This depends
upon two important lengths: the proton coherence length $\xi_p$
and the London penetration depth $\lambda_L$. If $\xi_p$ is larger
than $\sqrt{2}\lambda_L$, there are small normal regions
containing flux interspersed with field-free superconducting
regions (type I superconductivity). If instead $\xi_p$ is smaller
than $\sqrt{2}\lambda_L$, magnetic flux can penetrate the
superconductor without destroying the superconducting state.
Realistic estimations show that $\xi_p$ is much less than the
London penetration depth $\lambda_L$ for protons in the neutron
star interior and hence the proton superconductor is expected to
exhibit type II behavior.

However after recent observation of long period precession of
neutron stars it was pointed out in the literature, see for
example  Link,\cite{link} that the axis of precession of some
neutron stars may not be aligned with the axis of magnetic field.
As the rotation of neutron stars causes a lattice of quantized
vortices to form in the superfluid neutron state and it is
generally believed that the proton superfluid is a type II
superconductivity, which means that it supports a stable lattice
of magnetic flux tubes in the presence of magnetic field, Link
suggested that this two type of vortices might interact quite
strongly due to the fact that the axis of the rotation and the
axis of magnetic field are not aligned.

Due to this reason it was discussed\cite{kirk} that
superconductivity inside neutron stars in fact may be type I
because conventional picture of II type superconductivity follows
from the standard analysis when only a single proton field is
considered. In particular it has been shown the correlation length
or coherence length $\xi$ should be replaced by the actual size of
proton vortices
\begin{equation}
\label{Kirk} \bar{\delta}=\xi/\sqrt{\epsilon} \ .
\end{equation}

Since the parameter $\epsilon$ arises from the strong interaction
between proton and neutron superfluids, which is about $10^{-2}$,
the length (\ref{Kirk}) is much bigger than the usual proton
coherence length. Consequently the Ginzburg-Landau parameter to be
bigger than $1/\sqrt{2}$, which in its turn causes that the
superconductivity inside neutron stars to be type I.

Naturally the question arises whether the general relativistic
effects arising from the strong gravitational field of neutron
star can change the type of superconductivity or not.  As we have
shown here, due to the gravitational effect the coherence length
is also modified by factor of $N$ (the value of the parameter $N$
can reach $\approx0.7$ for the typical neutron stars), but it does
not lead to change of the type of superconductivity, as
gravitational field reduces the magnitude of the penetration depth
by the same factor so then the Ginzburg-Landau parameter will
remain unchanged.

It is well known that in superconductors of II type below lower
critical field $B_{c1}$ there will be complete expulsion of the
field; above an upper critical field $B_{c2}$ superconductivity
will be destroyed; in the intermediate range $B_{c1}<B<B_{c2}$ the
superconductor will allow the magnetic field to penetrate, not
homogeneously but confined to quantized flux tubes. Each fluxoid
carries a quantum of magnetic flux $\Phi_0 = hc/2e =2 \times
10^{-7}$ Gauss $\cdot$ cm$^2$. Within the core of a fluxoid (of
radius $\xi$), the matter is in its "normal" state. The field
strength rises to about $B_{c1}$ within this normal core. Around
this the matter is in the superconducting state, and the field
strength decreases exponentially away from the core with a scale
length $\lambda_L$. For protons in the interior of a neutron star,
the coherence length is $\xi_p=0.6 \times
10^{-12}\rho_{p,13}^{1/3}\Delta_p^{-1}$ cm and the London
penetration depth is
$\lambda_L=0.9\times10^{-11}\rho_{p,13}^{-1/2}$ cm , where
$\rho_{p,13}$ is the mass density of protons in units of $10^{13}$
g $\cdot$ cm$^{-3}$, and $\Delta_p$ is the energy gap of the
proton superconductor in MeV (Ref. 19).

The vortex filaments consisting of neutron vortices and proton
fluxoids inside superconducting region of neutron stars interact
each other with an interaction energy which is equivalent to
repulsion\cite{lifshitz}
\begin{equation}
\label{filament}
{\mathcal{E}}_{12}=\left(\frac{\Phi_0}{4\pi\lambda}\right)^2K_0\left(\frac{r}{\lambda}\right)
 \ .
 \end{equation}

 Due to the gravitational effect the interaction energy would be
 less when compared with what could be expected in the flat space-time by a factor
 \begin{equation}
 \label{factor}
 N^{-7/2}e^{\frac{d}{\lambda}\frac{N-1}{N}}
 \end{equation}
which is tiny and may be neglected during accounting one of the
main forces which has been acted on proton fluxoids.  Here
$d=5\times10^{-10}(B_{12})^{-1}$ is the spacing between proton
fluxoids. But there are the most important forces that act on
proton fluxoids such as magnus force, buoyancy force, drag force
and tension of fluxoids. But there are the most important forces
that act on proton fluxoids such as magnus force, buoyancy force,
drag force and tension of fluxoids. The collective effect of these
forces leads fluxoids to move radially outward. The
force\cite{ding}
 \begin{equation}
 \label{magnus}
 f_n=\frac{2\Phi_0\rho_cR_c\Omega_c(\Omega_c-\Omega_s)}{B_c} \ ,
 \end{equation}
acting on unit length of fluxoid (due to neutron vortices) does
not depend on gravitational  field and remains unchanged. Here
$R_c$ is the core radius, $\Omega_s$ and $\Omega_c$ are the
rotation rate of the core superfluid and the rotation rate of the
crust respectively. However the other forces such as buoyancy
force arising from the existence of the magnetic stress in the
core of a fluxoid and drug force that acted by the electron gas to
a flux tube moving with the velocity $\mathbf{v}$ depend on the
gravitational field and would be modified when compared with the
Newtonian ones\cite{harvey}
 \begin{eqnarray}
 \label{buoyant}
f_b=\left(\frac{\Phi_0}{4\pi\lambda
N}\right)^2\frac{1}{R_c}ln\left(\frac{\lambda}{\xi}\right) \ ,
\\ \nonumber\\
\label{drag}
f_d=-\frac{3\pi}{64}\frac{n_ee^2\Phi_0^2}{E_f(e)\lambda
N^2}\frac{v_p}{c} \ ,
\end{eqnarray}
where  $\tilde{E}_f=NE_f(e)$ is the electron Fermi energy in the
stationary gravitational field. In spite of the fact the
gravitational field of neutron stars does not change the value of
these forces acted on fluxoids enough, the motion of fluxoids
inside neutron star might be changed. Considering the equation of
motion of fluxoids\cite{ding} one can eventually see that
difference of the rotation rate of the core superfluid from the
rotation rate of the crust also depends on the value of the
gravitational field.

 Since the magnetic field decay from the core of neutron stars depends
 on interpinning of neutron superfluid and proton superconducting
 fluid\cite{ding}, then the slightly modified equation of motion
 leads to the change of the decay timescale, but this modification is
 very small, and could not be important for considering
the lifetime of the magnetic field persisting in neutron stars.


\section{Conclusion}

Following up our previous research on the electromagnetic
properties of superconductors in gravitational field we have
considered some general-relativistic effects associated with the
influence of stationary gravitational field on quasi-stationary
electromagnetic effects in conductors and superconductors.

By solving Maxwell equations for the time-dependent magnetic field
penetrating inside the conductor we have shown that the skin depth
in the conductors embedded in the gravitational field strongly
depends on the gravitational field. However this effect is almost
negligible for the laboratory conductors in the weak gravitational
field of the Earth.

The penetration depth for the magnetic field and the coherence
length in superconductors also depend on the gravitational field
and proportional to the lapse function of the gravitational
object. However the parameter of Ginzburg-Landau remains unchanged
and  does not lead to the general relativistic modification of the
type of the superconductivity.

Two critical magnetic fields that destroy the superconducting
state also depend on the gravitational field: the gravitational
field increases the value of the critical magnetic field. Thus it
has been shown that the strong gravitational field does effect to
the quasi-stationary electromagnetic processes inside
superconductors and one must take them into account during
studying neutron star physics.

\section*{Acknowledgments}

Authors thank TWAS and AS-ICTP for the support towards their
travel expenses to India and IUCAA for warm hospitality during
their stay in Pune. B. Ahmedov acknowledges the partial support by
the NATO Reintegration Grant EAP.RIG.981259. This research is also
supported in part by the Uzbekistan Foundation for Fundamental
Research (project 02-04) and projects F2.1.09 and F2.2.06 of the
Uzbekistan Centre of Science and Technology.


\end{document}